\documentclass[conference]{IEEEtran}
\IEEEoverridecommandlockouts

\usepackage{ifpdf}
\usepackage{cite}

\ifCLASSINFOpdf
  \usepackage[pdftex]{graphicx}
\else
  \usepackage[dvips]{graphicx}
\fi

\usepackage{amsmath}

\usepackage{algorithm,algorithmic}

\usepackage{array}

\ifCLASSOPTIONcompsoc
 \usepackage[caption=false,font=normalsize,labelfont=sf,textfont=sf]{subfig}
\else
 \usepackage[caption=false,font=footnotesize]{subfig}
\fi

\usepackage{fixltx2e}

\usepackage{stfloats}

\usepackage{url}

\usepackage{booktabs}
\usepackage{multirow}
\usepackage{authblk}

\begin{document}

\title{A Method for Efficient Heterogeneous Parallel Compilation: A Cryptography Case Study}

\author[1,2]{Zhiyuan Tan}
\author[1,2]{Liutong Han}
\author[1]{Mingjie Xing}
\author[1]{Yanjun Wu\textsuperscript{\textdagger}\thanks{\textsuperscript{\textdagger}Corresponding author: Yanjun Wu (yanjun@iscas.ac.cn)}}

\affil[1]{Intelligent Software Research Center, Institute of Software, Chinese Academy of Sciences, Beijing, China}
\affil[2]{University of Chinese Academy of Sciences, Beijing, China}

\maketitle

\begin{abstract}
In the era of diminishing returns from Moore’s Law, heterogeneous computing systems have emerged as a vital approach to enhance computational efficiency. This paper introduces a novel MLIR-based dialect, named \textit{hyper}, designed to optimize data management and parallel computation across diverse hardware architectures. The \textit{hyper} dialect abstracts the complexities of heterogeneous computing by providing a unified compilation framework that efficiently schedules tasks and manages data communication. To demonstrate its capabilities, we present \textit{HETOCompiler}, a cryptography-focused compiler prototype that implements multiple hash algorithms and enables their execution on heterogeneous systems. The proposed approach achieves performance improvements over existing programming models for heterogeneous computing (OpenCL), offering an average speedup of 1.93x, 1.18x, and 1.12x for SHA-1, MD5, and SM3 algorithms, respectively. Our findings highlight the potential of the \textit{hyper} dialect in harnessing the full computational power of heterogeneous devices, advancing the field of compiler design for heterogeneous systems.
\end{abstract}

\section{Introduction}

As transistor sizes on silicon chips approach physical limits, the efficacy of Moore's Law \cite{Moore1965} has progressively diminished. To overcome the physical constraints hindering the enhancement of general-purpose processor performance, considerable exploration has been devoted to hardware architecture advancements. Among these endeavors, heterogeneous computing systems have emerged as a highly promising solution. Heterogeneous computing systems encompass complex architectures comprising diverse processors, accelerators, and other computing devices. Typically, these systems integrate a general-purpose processor (e.g., CPU) alongside multiple specialized accelerators (e.g., GPUs, FPGAs, and CGRAs), which provide a critical technological means for the vision of ubiquitous computing.

\begin{figure}[t]
\centering
\includegraphics[width=0.5\textwidth]{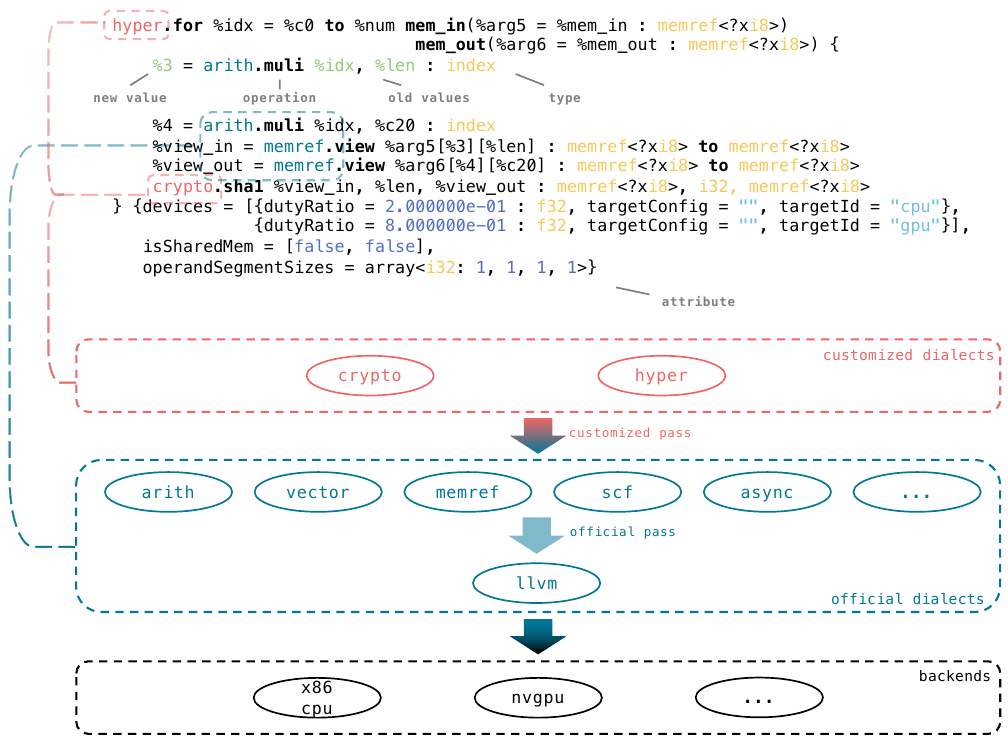}
\caption{An MLIR example with multiple coexisting dialects describing the core semantics of the SHA-1 algorithm on different heterogeneous devices.}
\label{fig:mlir-example}
\end{figure}

To effectively harness the diverse computing resources of general heterogeneous computing systems for large-scale computations, the modern compilers should have the ability to compile the source code targeting various heterogeneous devices and the potential to split the tasks for each backend. Multi-Level Intermediate Representation (MLIR) \cite{mlir} is a rising foundational infrastructure with the power to abstract the backends on a higher level above LLVM IR. It introduces the pivotal concept of \textit{dialects}, enabling the abstraction of structures at various levels. Users can describe algorithmic computation patterns or specific hardware architecture models by seamlessly integrating different levels of IR inside different dialects, which will be progressively lowered to a unified LLVM IR and then be translated into hardware specific machine code ultimately (as shown in Fig.~\ref{fig:mlir-example}). MLIR upstream offers some backend specific dialects like '\textit{arm\_neon}', '\textit{x86vector}', '\textit{amdgpu}', '\textit{nvgpu}' etc., nevertheless it lacks the semantics to define general behaviors of miscellaneous devices on a heterogeneous computing system. This absence may bring challenges in compiler design for heterogeneous computing systems, as various compilation passes should be considered for different hardwares.

To bridge this gap, this paper introduces a meticulously designed MLIR dialect, referred to as the \textit{hyper} dialect, which abstracts both data management and parallel computation functionalities for heterogeneous devices. This dialect can be further transformed into low-level, target-specific dialects to describe specialized behaviors on specific hardware. To demonstrate the capabilities of this dialect, we focus on a cryptographic scenario involving extensive parallel hash computations over large volumes of input messages. Consequently, we developed \textit{HETOCompiler}, an end-to-end cryptography compiler prototype built upon MLIR. Utilizing the concept of \textit{dialects} within MLIR, the compiler implements three commonly used hash encryption algorithms and enables their parallel execution on heterogeneous devices through our \textit{hyper} dialect. The main contributions of this paper are as follows:

\begin{itemize}
\item We have designed a generic heterogeneous scheduling dialect named \textit{hyper}. This dialect abstracts the intricacies of data communication between host processors and heterogeneous devices, as well as the task splitting and scheduling on heterogeneous platforms, filling the gap in upstream MLIR.
\item We introduce a compiler prototype for cryptographic computation scenario, which orchestrates
an end-to-end process gradually transitioning the source code from high-level hash algorithms to device-specific executables. This comprehensive approach enhances the efficiency of hash encryption algorithms across heterogeneous computing systems.
\item The results of our experiment show that compared with the commonly used heterogeneous programming model OpenCL, our compiler achieves an average of 1.93x, 1.18x and 1.12x performance boost for SHA1, MD5 and SM3 algorithms respectively.
\end{itemize}

The organizational structure of this paper is as follows: Sect.~\ref{sec:related-work} discusses existing related work. Sect.~\ref{sec:hyper-dialect} introduces the designed semantic rules and the optimization techniques inside \textit{hyper} dialect. Sect.~\ref{sec:heto-compiler} elaborates on the main architecture and design details of \textit{HETOCompiler}. Sect.~\ref{sec:experiments-results} showcases the experimental results of \textit{HETOCompiler}'s compilation outputs on different heterogeneous systems. Sect.~\ref{sec:conclusion} summarizes the paper and proposes prospects for future work.

\begin{figure*}[t]
\centering
\includegraphics[width=0.8\textwidth]{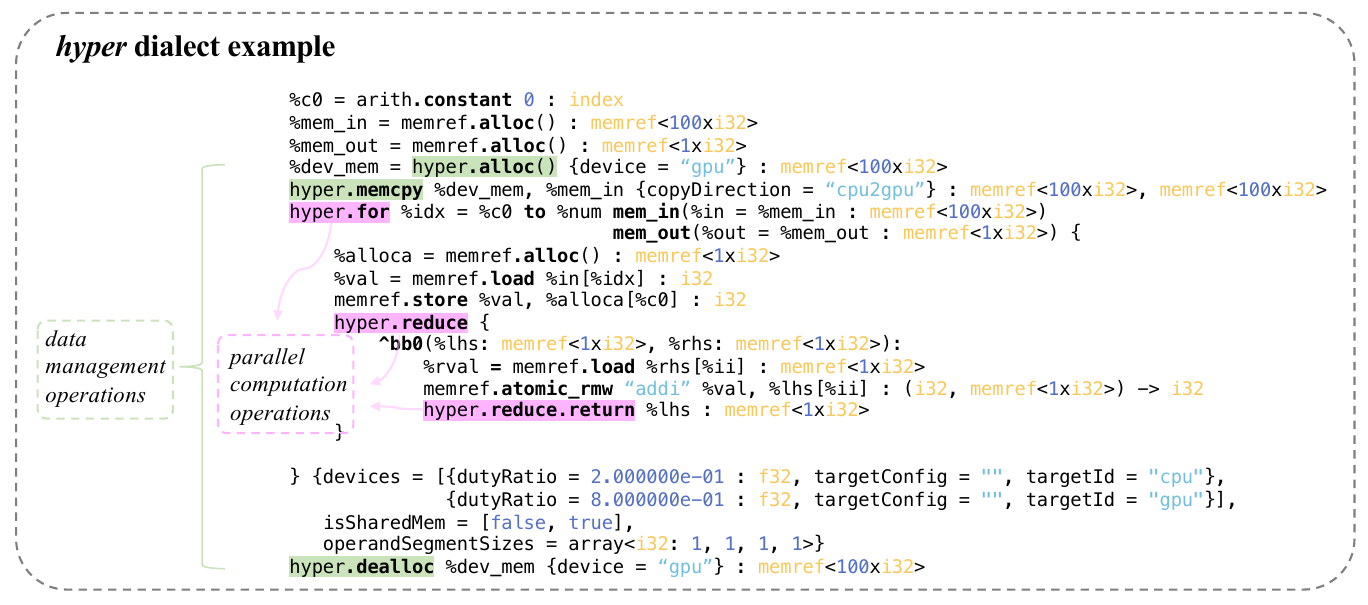}
\caption{A simple example program which sums an array using \textit{hyper} dialect written in MLIR.}
\label{fig:hyper-operations}
\end{figure*}

\section{Related Work}
\label{sec:related-work}
The programming of heterogeneous systems has been a significant research area due to its potential for energy-efficient, high-performance computing. Fang \textit{et al.} \cite{fang2020parallel} provide a comprehensive survey of parallel programming models for heterogeneous many-core architectures, emphasizing both low-level and high-level programming models. Low-level models, such as CUDA \cite{CUDA} and OpenCL \cite{OpenCL}, are closely tied to specific hardware and offer fine-grained control but require extensive knowledge of the underlying architecture. In contrast, high-level models like SYCL \cite{SYCL} and directive-based models such as OpenMP \cite{chandra2001parallel} and OpenACC \cite{OpenACC} aim to enhance programmability by abstracting hardware details, thereby making it easier for developers to write portable and efficient code for different platforms. Also, there are vendor-specific programming models like Intel’s OneAPI \cite{oneAPI} aims to provide a single programming model that supports CPUs, GPUs, and FPGAs, promoting code portability across different hardware architectures. Unlike some specialized programming models, our approach seeks to offer a compilation technique capable of distributing different tasks across various devices, rather than being tailored to specific hardware from particular vendors.

In the realm of cryptography-specific compilers and languages, \texttt{CryptOpt} \cite{kuepper2023cryptopt} stands out as a pioneering compilation pipeline excelling in transforming cryptography programs into assembly code that outperforms outputs generated by GCC or Clang compilers. Moreover, the \texttt{Jasmin} programming language \cite{almeida2017jasmin} stands out for offering an efficient mechanism to generate cryptographic programs characterized by predictably high-speed and high-assurance attributes. On the other hand, \texttt{Usuba} \cite{mercadier2019usuba} and \texttt{EasyBC} \cite{sun2024easybc} are cryptographic domain-specific languages (DSLs) specifically tailored for block ciphers, emphasizing optimized performance in such contexts. Another notable example is \texttt{Vale} \cite{bond2017vale}, a novel DSL tailored for verifying manually tuned high-performance cryptographic assembly code by inspecting the converted AST derived from annotated assembly code. Additionally, \texttt{FaCT} \cite{fact2019}, yet another cryptographic DSL equipped with a secrecy type system, facilitates the translation of high-level code into constant-time LLVM bitcode, thereby leveraging LLVM back-ends for code generation as well as our prototype compiler. While these works excel in various aspects of cryptography and demonstrate exceptional performance, they primarily focus on optimizing single-message operations. In contrast, our research centers on heterogeneous compilation techniques, with cryptography serving merely as a case study. Our analysis will illustrate that, in certain scenarios, traditional cryptographic solutions may be inadequate, necessitating the utilization of the full computational capabilities offered by all available heterogeneous devices within the computing environment.

\begin{figure*}[t]
\centering
\includegraphics[width=0.8\textwidth]{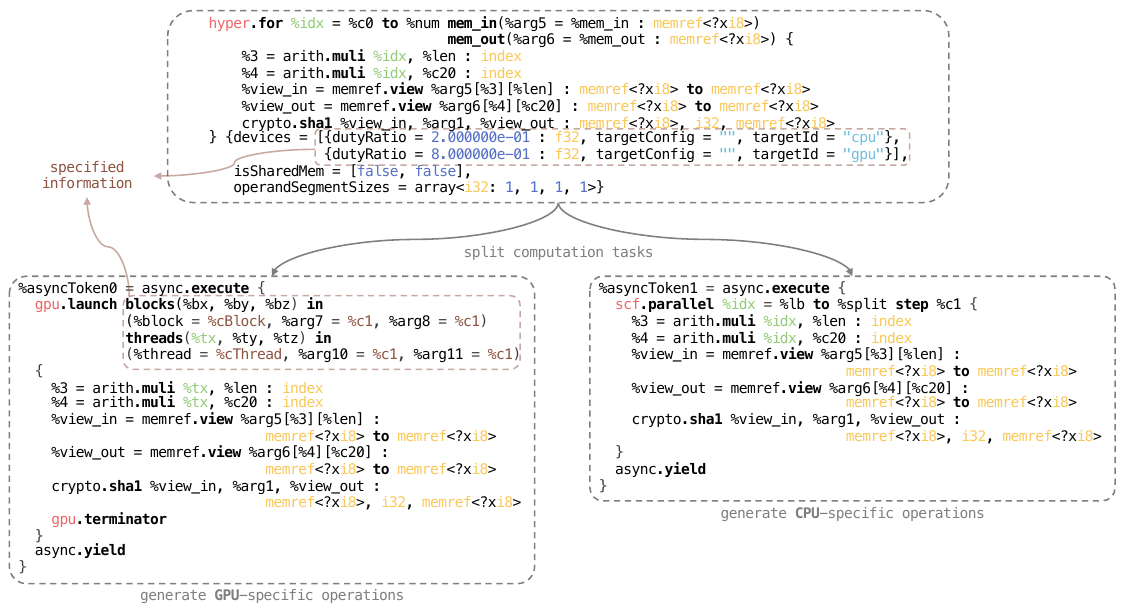}
\caption{A transformation example from hyper.\textbf{for} operation to target-specific operations.}
\label{fig:hyper-for-example}
\end{figure*}

\section{Proposed MLIR Dialect}
\label{sec:hyper-dialect}
In MLIR, a dialect encompasses fundamental elements such as operations, attributes, and types, delineating the syntax and semantic rules pertinent to that domain (as decipited in Fig.~\ref{fig:mlir-example}). Operations, being the fundamental units, articulate program semantics in MLIR. They preserve the static single assignment (SSA) property by consuming existing values and yielding new ones. Attributes furnish pertinent information accessible at compile time, thereby enabling optimizations based on these information. Types encapsulate compile-time information about values, thereby upholding MLIR's internal type system.

In \textit{hyper} dialect, we encapsulate the processes of storage management on heterogeneous devices, the data exchange between the host and these devices, and the scheduling of computational tasks across diverse hardware components. This dialect serves as a comprehensive abstraction layer, encompassing operators that abstract both data management and parallel computation functionalities of heterogeneous devices. Furthermore, by employing specialized transformations, we fine-tune storage utilization and task execution efficiency across multiple heterogeneous devices.

\subsection{Hyper Operations}
Operations of \textit{hyper} dialect serve as abstractions for two fundamental aspects: data management and parallel computation. Figure~\ref{fig:hyper-operations} shows an example program that calculates the sum of all elements in an array, by splitting the computation task into two different devices and reducing the collected results using our \textit{hyper} dialect. Although this dialect is not initially designed for this simple scenario, it is adequate to demonstrate all the operations inside \textit{hyper} dialect.

Regarding data management, the \texttt{hyper.\textbf{alloc}} operation elucidates the semantics indicating the allocation of storage space on the computational device, with the \texttt{device} attribute precisely delineating the target device. Conversely, the \texttt{hyper.\textbf{dealloc}} operation encapsulates semantics that are antithetical to those of \texttt{hyper.\textbf{alloc}}, denoting the release of allocated storage space on the corresponding device. Furthermore, the \texttt{hyper.\textbf{memcpy}} operation describes the data transfer behaviors between the host system and the heterogeneous devices.

For parallel computation, we introduce a fundamental operation, denoted as \texttt{hyper.\textbf{for}}, which encapsulates the logic for task partitioning and execution across various devices. As depicted in Figure~\ref{fig:hyper-operations}, the \texttt{hyper.\textbf{for}} operator delineates loop boundaries by lower and upper bounds, employing a fixed step size of one. Additionally, it accommodates optional parameters, \texttt{mem\_in} for storing input elements of the computational task, and \texttt{mem\_out} for the output. The subsequent region of the \texttt{hyper.\textbf{for}} operation describes the specific computation logic tailored for parallel tasks, thereby applicable across all heterogeneous devices. Furthermore, the \texttt{devices} attribute lists out the available devices, where \texttt{dutyRatio} denotes the workload distribution ratio of the computation task on each device, \texttt{targetConfig} specifies device-specific configurations, and \texttt{targetId} denotes the type and identifier of the available devices. Moreover, the \texttt{isSharedMem} attribute signifies whether input/output partitioning is requisite, thus satisfying the diverse requirements of various computational scenarios.

To augment the versatility of the \texttt{hyper.\textbf{for}} operation, we introduce the \texttt{hyper.\textbf{reduce}} operation designed for reduction operations on computation outcomes, accompanied by the \texttt{hyper.\textbf{reduce.return}} operation, serving as the terminator of its region. The optional \texttt{hyper.\textbf{reduce}} operation must be a sub-operation of \texttt{hyper.\textbf{for}}, and there is a basic block attached in its region. This basic block elaborates the logic for reducing disparate computation outcomes by taking two results with the same type from different threads/devices as its parameters. The atomic operations within the basic block (like \texttt{memref.\textbf{atomic\_rmw}} in Figure~\ref{fig:hyper-operations}) is a must to ensure the accuracy of the results. Reductions are executed among diverse threads on the same device and across heterogeneous devices, culminating in a final reduction outcome.

\subsection{Hyper Transformations}
For operations responsible for data management, we convert them into device-specific data management operators. For instance, when targeting GPUs, the \texttt{hyper.\textbf{alloc}}, \texttt{hyper.\textbf{dealloc}}, and \texttt{hyper.\textbf{memcpy}} operations undergo transformation into \texttt{gpu.\textbf{alloc}}, \texttt{gpu.\textbf{dealloc}}, and \texttt{gpu.\textbf{memcpy}} operations, respectively, delineating behaviors specific to GPU devices. It is worth noting that such transformation may introduce additional overhead in terms of space allocation and data copying on the host CPU. To address this concern, we have devised an optimization pass grounded on mapping rules aimed at mitigating unnecessary data management overhead on the host CPU.

\begin{algorithm}
 \label{alg:mem-opt}
 \caption{Host Memory Optimization Process}
 \begin{algorithmic}[1]
 \renewcommand{\algorithmicrequire}{\textbf{Input:}}
 \renewcommand{\algorithmicensure}{\textbf{Output:}}
 \renewcommand{\algorithmicforall}{\textbf{foreach}}
 \REQUIRE Node list $N_u$ and $N_v$; Edge list $E$
 \ENSURE  Node list $N'_v$, Edge list $E'$
 \STATE \textit{Initialisation}: $N'_v \gets \{N_v, N_u\}$, $E' \gets E$
  \FORALL{node $u \in N_u$}
   \FORALL{edge $(u, v) \in E$}
    \IF{$\exists$ edge $(w, u)$ such that $w \neq v$}
     \STATE $E' \gets (w, v)$
     \STATE $remove\_node(u)$
     \STATE $remove\_edge((w, u))$
    \ELSE
     \STATE $replace\_node(u, v)$
    \ENDIF
    \STATE $remove\_edge((u, v))$
   \ENDFOR
   \FORALL{edge $(v, u) \in E$}
    \STATE $replace\_node(u, v)$
    \STATE $remove\_edge((v, u))$
   \ENDFOR
   \IF{$exist(u)$}
     \STATE create new node $n$
     \STATE $replace\_node(u, n)$
   \ENDIF
  \ENDFOR
 \RETURN $N'_v$, $E'$ 
 \end{algorithmic} 
\end{algorithm}

Memory allocation operations can be represented as the creation of nodes in a directed graph (DG), while the \texttt{hyper.\textbf{memcpy}} operations denote the direction of data flow, corresponding to the edges in the graph. By scanning all IRs, we can construct a DG to model the overall data management. We define two node lists: $N_u$, which includes all nodes corresponding to \texttt{hyper.\textbf{alloc}} operations, and $N_v$, which includes nodes corresponding to \texttt{memref.\textbf{alloc}} operations. Given that \texttt{hyper.\textbf{memcpy}} operations describe data transfers between host and device, the involved memory locations must originate from distinct sources. Thus, the edge list E is formed based on all \texttt{hyper.\textbf{memcpy}} operations, connecting nodes from the two lists, $N_u$ and $N_v$. With this representation, we propose a graph-based optimization algorithm as outlined in Algorithm 1. The core idea of this algorithm is to utilize the edges in the graph to establish a mapping relationship between host and device memory. By replacing device memory nodes with corresponding host memory nodes, assuming all memory resides on the host, we can effectively reduce redundant memory overhead by transforming the simplified DG back to IRs.

\begin{figure}[h]
\centering
\includegraphics[width=0.5\textwidth]{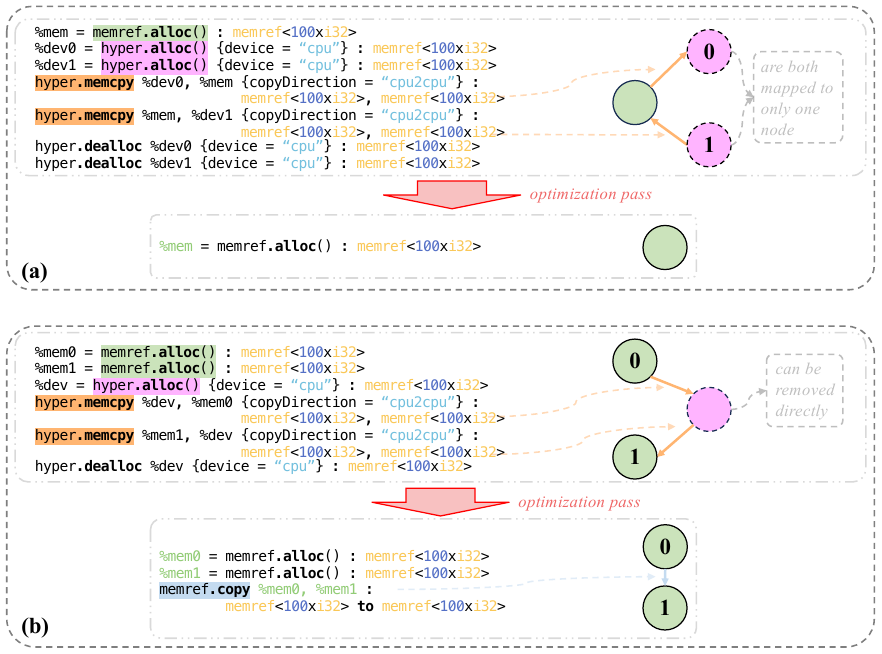}
\caption{Illustrative examples demonstrating how the graph-based algorithm optimizes redundant data management overhead on the host CPU.}
\label{fig:hyper-host-mem-opt}
\end{figure}

Figure~\ref{fig:hyper-host-mem-opt} provides illustrative examples of the proposed algorithm’s working process. In Figure~\ref{fig:hyper-host-mem-opt}(a), both device nodes 0 and 1 are mapped to a single host node, indicating that the associated \textit{hyper} operations can be eliminated, resulting in only the host memory allocation remaining. In contrast, Figure~\ref{fig:hyper-host-mem-opt}(b) depicts a scenario where the device node is mapped to two distinct host nodes, establishing a data flow from host node 0 to host node 1. Here, it is necessary to remove the device node and introduce an additional \texttt{memref.\textbf{copy}} operation to facilitate the data transfer. The optimization pass iteratively identifies and optimizes specific graph patterns until no device nodes remain in the directed graph.

For the \texttt{hyper.\textbf{for}} operation, which describes parallel computing semantics across heterogeneous devices, dedicated transformation passes must be devised to convert it into intermediate representations for distinct backend architectures. To elucidate our transformation methodology, we employ the code snippet depicted in Figure~\ref{fig:mlir-example} and showcase the transformation of the \texttt{hyper.\textbf{for}} operator for CPU and GPU backends, as exemplified in Figure~\ref{fig:hyper-for-example}. Initially, this pass computes and partitions the input and output spaces for each backend based on the respective workload ratios. Subsequently, for the GPU backend, the \texttt{hyper.\textbf{for}} operation undergoes transformation into the \texttt{gpu.\textbf{launch}} operation within the \textit{gpu} dialect to initialize the kernel. Conversely, for the CPU backend, it generates the \texttt{scf.\textbf{parallel}} operation within the MLIR core \textit{scf} dialect, containing the parallel loop computation process using the multi-core capability of the CPU. It is crucial to note that the iteration variables necessitate appropriate mapping within distinct loop bodies to ensure computational correctness. The determination of workload ratios and kernel launch parameters depicted in Figure~\ref{fig:hyper-for-example} can be set freely by users, underscoring that this dialect supports an external module for heterogeneous device workload scheduling algorithms. However, the investigation into heterogeneous device workload scheduling algorithms lies beyond the scope of this paper and represents one of our future research goals.

\begin{figure*}[t]
\centering
\includegraphics[width=0.95\textwidth]{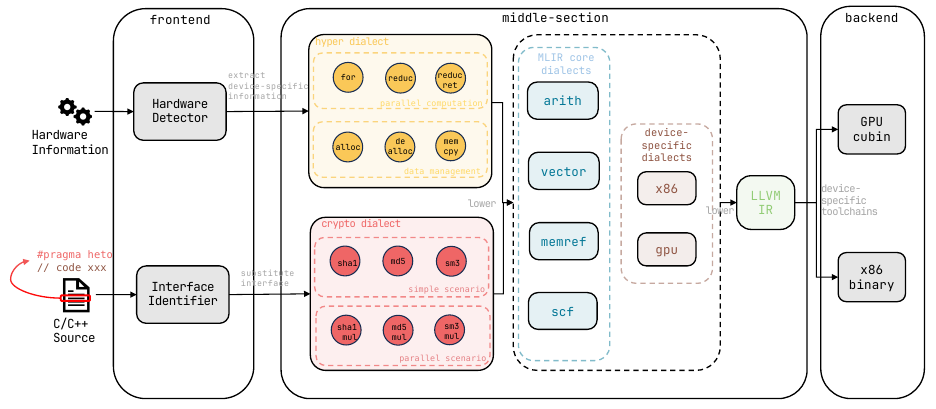}
\caption{The overview structure of \textit{HETOCompiler}.}
\label{fig:compiler-overview}
\end{figure*}

\section{HETOCompiler}
\label{sec:heto-compiler}

To demonstrate the capabilities of the proposed dialect, we focus on a cryptographic scenario characterized by extensive parallel hash computations for large volumes of input messages. In this context, we proposed \textit{HETOCompiler}, an end-to-end cryptography compiler prototype based on MLIR. By utilizing MLIR’s dialect framework, the compiler implements three widely-used hash encryption algorithms, enabling parallel execution on heterogeneous devices through our custom hyper dialect. Figure~\ref{fig:compiler-overview} illustrates the overarching design of \textit{HETOCompiler}, which primarily comprises three core components: frontend, middle-section, and backend.

\subsection{Compiler Frontend}
\label{subsec:compiler-frontend}
The compiler frontend operates as the "eyes" of the compilation system, responsible for gathering external information. Primarily, our frontend comprises two key components: the \textbf{interface recognizer} and the \textbf{hardware detector}, as illustrated in Figure~\ref{fig:compiler-overview}.
The former assumes the pivotal role of identifying the pertinent interfaces provided by us within C/C++ source code, while the latter possesses the capability to automatically detect hardware devices on the computing system.

\subsubsection{Interface Identifier}
Our frontend supports the recognition and compilation of C/C++ source code. Users can leverage the cryptographic-related interfaces provided by our compiler and assign special tags (\texttt{\#pragma}-like) to these interfaces when constructing their applications. Upon parsing the abstract syntax tree (AST) derived from the source code, the interface recognizer identifies the designated special tags (\texttt{\#pragma heto} in Figure~\ref{fig:compiler-overview}). It subsequently translates the user-invoked interfaces into corresponding cryptographic dialect operators within the AST. These cryptographic dialect operators undergo specific transformations via our predefined conversion and lowering passes. Meanwhile, the remaining segments of the AST are directly utilized by LLVM's existing C/C++ backend for further optimization and compilation. Ultimately, these components are combined to yield our application specifically optimized for cryptographic algorithms.

\subsubsection{Hardware Detector}
The hardware detector of the frontend is capable of acquiring and parsing hardware-specific information during the build process, subsequently storing it as a singleton object for utilization in subsequent compilation processes. This object includes fields such as the number and model of devices, along with partitioning ratios of tasks on diverse devices. Such information holds paramount importance for task partitioning and hardware parameter configuration within the heterogeneous scheduling \textit{hyper} dialect. For instance, within the frontend, we maintain a table of NVIDIA GPU information, which includes details such as compute capability, maximum block size, and maximum grid size for specific GPU models, as depicted in Table~\ref{tab:exp-env}. The max block/grid size information aids the \textit{hyper} dialect in determining the appropriate number of threads to be configured when launching GPU kernels.

\subsection{Compiler Middle-Section}
\label{subsec:middle-section}
The middle-section is the core part of the compiler dedicated to optimization. It encompasses our customized dialects alongside the core dialects offered by MLIR, in addition to passes of dialect conversion or lowering. Leveraging the robust multi-level semantic representation capability of MLIR, we can progressively lower our custom dialects to intermediate representations with hardware-specific semantics. Throughout this process, targeted optimizations are applied to obtain the optimal LLVM IR.

To address the needs of the cryptographic domain, we have developed and integrated a dedicated \textit{crypto} dialect into the intermediate layer of our prototype compiler. The main aim of this dialect is to provide a device-agnostic, general-purpose implementation of common hash algorithms that can be run on any device with the same intermediate representations. It implements two categories of algorithm-level operations for three widely-used hash algorithms (SHA-1, MD5, and SM3): one for simple scenarios, where the digest is computed for a single message, and another for parallel scenarios, where the digests are computed for multiple messages simultaneously. It is noteworthy that some CPUs in the x86 architecture support the SHA-NI extension instruction set\cite{gulley2013intel}, which provides hardware acceleration for SHA algorithms. To leverage these hardware capabilities fully, we have implemented a dedicated pass that translates the \texttt{crypto.\textbf{sha1}} operation into SHA-NI instructions. This pass can be automatically applied when the specific hardware is detected by the hardware detector or manually disabled by the user if desired.

In parallel scenarios, the \textit{crypto} dialect and the previously mentioned \textit{hyper} dialect collaborate to define the semantics for computing a large number of digests across various heterogeneous devices concurrently. The \texttt{hyper.for} operation determines the task partitioning and GPU thread configuration based on the hardware information provided by the hardware detector. Following a series of transformations on these dialects, tasks with general representations of hash algorithms can be effectively scheduled across different devices, utilizing a combination of operations from the standard MLIR dialects to construct the IR. During this phase, optimization passes provided by the upstream MLIR can be applied to produce an optimal LLVM IR.

\subsection{Compiler Backend}
\label{subsec:compiler-backend}
The backend represents a critical component of the compiler concerning code generation. By employing LLVM IR as a hardware-independent unified intermediate representation, we can ultimately lower various MLIR dialects to robust, hardware-agnostic LLVM IR. Since multiple hardware vendors typically provide passes for translating LLVM IR into hardware-specific machine code, existing hardware backends for code generation can be effectively utilized. For the x86 architecture, LLVM natively supports the compilation of LLVM IR into executable or object files, leveraging the \textit{Clang} driver for x86 backend code generation. Conversely, for NVIDIA GPUs, due to the distinct compilation and linking requirements for host and device code, we utilize the \textit{NVCC} toolchain provided by NVIDIA for backend code generation on NVIDIA GPUs. These backend toolchains subsequently perform tailored optimizations to produce optimal backend-specific executable files.

\section{Experiments \& Results}
\label{sec:experiments-results}
\subsection{Experimental Setup}
\subsubsection{Environments}
To evaluate the effectiveness of our approach across different architectures, we deployed the \textit{HETOCompiler} on two servers with distinct configurations. The detailed specifications of the experimental environment are provided in Table~\ref{tab:exp-env}. Each server is equipped with a multi-core x86 CPU and an NVIDIA GPU. Notably, the CPU in Server B is based on the Intel Skylake microarchitecture, which lacks the support of the SHA-NI cryptographic acceleration instructions. These two platform configurations demonstrate the versatility of our approach under different hardware conditions.

\begin{table}[h]
    \centering
    \caption{Experimental environments for our evaluation.}
    \label{tab:exp-env}
    \begin{tabular}{c c c}
        \toprule
         & Server A & Server B \\
        \midrule
        Operating System & Ubuntu 22.04.2 LTS & Ubuntu 22.04.3 LTS\\
        Host Mem & 126 GB & 188 GB\\
        \multirow{2}{*}{CPU} & 13th Gen Intel(R) & Intel(R) Xeon(R)\\
         & Core(TM) i7-13700K & Gold 5218R\\
        CPU Cores & 24 & 80\\
        SHA-NI Support & Yes & No\\
        \multirow{2}{*}{GPU} & GeForce RTX 2060 & \multirow{2}{*}{GeForce RTX 3090}\\
         & SUPER & \\
        GPU Mem & 8 GB & 24 GB\\
        Compute Capability & \multirow{2}{*}{sm\_75} & \multirow{2}{*}{sm\_86}\\
        (GPU) &  & \\
        Max Grid Size & \multirow{2}{*}{$2^{31}-1$} & \multirow{2}{*}{$2^{31}-1$}\\
        (x dimension) &  & \\
        Max Block Size & \multirow{2}{*}{$65536$} & \multirow{2}{*}{$65536$}\\
        (x dimension) &  & \\
        
        \bottomrule
    \end{tabular}
\end{table}

\subsubsection{Data Preparation}
To create a large-scale data scenario that is more efficiently processed using multiple devices in a heterogeneous system, we set up a straightforward scenario focused on massive hash values mapping. In this scenario, extensive computation is required to generate hash values for short messages, as reversing a hash value to obtain the original long, complex message is nearly impossible. Specifically, we computed the SHA-1, MD5, and SM3 hash values for various messages consisting of a 9-character length using the character set {0-9}. The total data processed for computing these hash values amounts to 27.01 GB for SHA-1, 23.28 GB for MD5, and 38.18 GB for SM3.

\subsubsection{Baseline Selection}
In order to demonstrate the limitations of high-performance cryptographic algorithm libraries in our experimental scenario, we have chosen to compare our prototype compiler against these libraries. Sodium \cite{libsodium}, a widely used cryptographic library, does not support the aforementioned hash algorithms. On the other hand, OpenSSL \cite{openssl} is a well-established cryptographic library known for its extensive support and optimization of a wide range of algorithms through the use of assembly language. As such, it serves as an appropriate baseline for our evaluation. For this study, we will use the latest stable release version v3.3.0 of OpenSSL. To ensure a fair comparison, we will employ OpenMP \cite{chandra2001parallel} to utilize all available CPU threads for massive message processing on the CPU when using OpenSSL.

Furthermore, as an optimization strategy for heterogeneous computing at the compilation level, it is essential to compare our approach with existing programming models designed for heterogeneous computing. OpenCL \cite{OpenCL} is an open standard for parallel programming of heterogeneous systems, enabling developers to harness the computational capabilities of various device types in a unified manner. We developed hand-optimized OpenCL kernels for these algorithms, and our MLIR implementations were informed by the OpenCL versions to ensure the generation of similar kernels. All programs were compiled using \textit{Clang 17.0.6} with the “-O3” optimization level, and all performance measurements were conducted using Google Benchmark \cite{googlebenchmark}.

\subsection{Comparison with OpenSSL}
\label{subsec:comp-openssl}
\begin{figure}[h]
\centering
\includegraphics[width=0.5\textwidth]{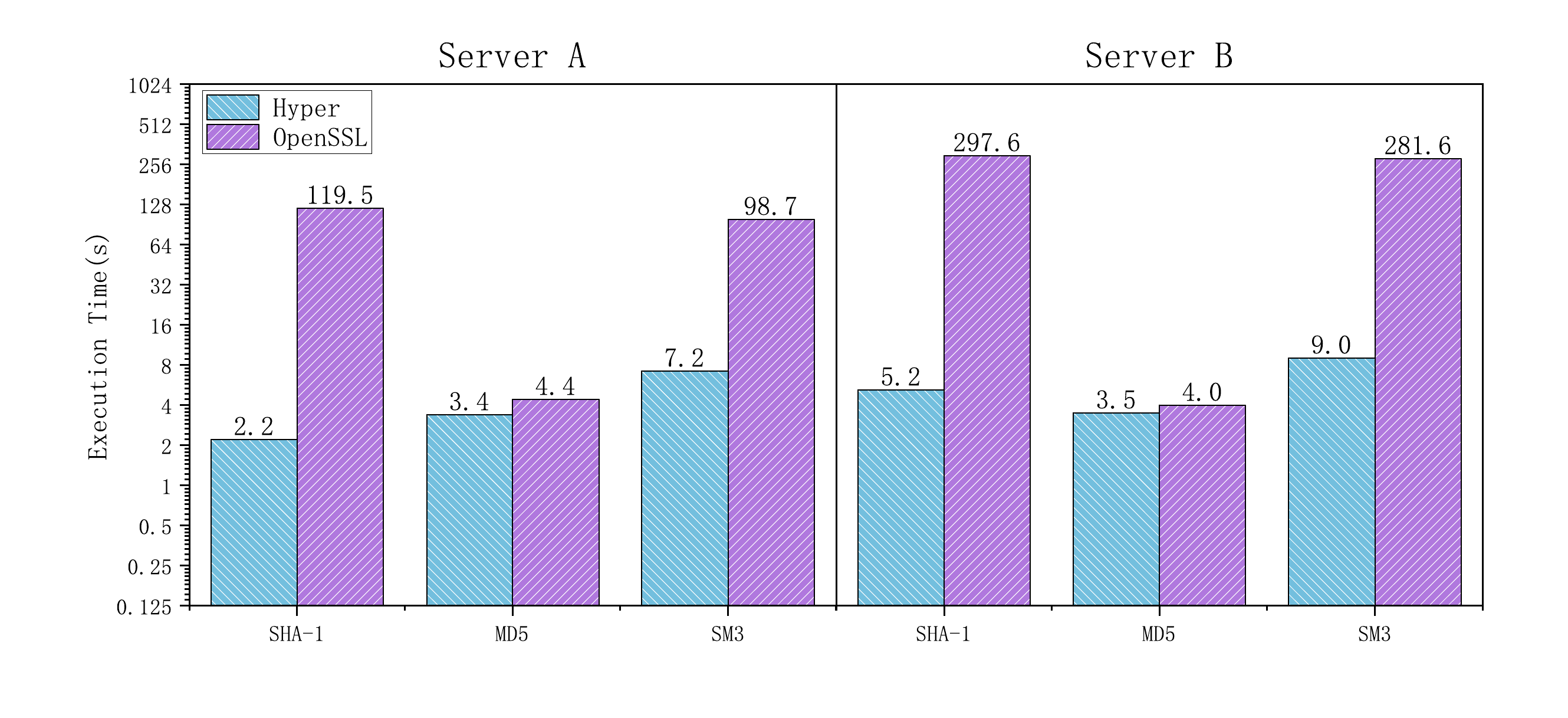}
\caption{Comparison results with OpenSSL on different servers.}
\label{fig:compWithSSL}
\end{figure}

To demonstrate that our computing scenario is not well-suited for traditional cryptographic libraries, we compared the performance of algorithms implemented in our prototype compiler with those in OpenSSL. The results, shown in Figure~\ref{fig:compWithSSL}, indicate average performance improvements of 55.8x for SHA-1, 1.2x for MD5, and 22.5x for SM3. It is important to note that the results of our approach were obtained by iteratively exploring various workload ratios within the same intervals to identify the optimal configuration. However, this optimization process is not claimed as a primary contribution of our work, as we only adopt the simplest policy. A detailed analysis of the workload scheduling policy will be provided in Sect. \ref{subsec:res-analysis}.

As shown in Figure~\ref{fig:compWithSSL}, substantial performance improvements can be observed for the SHA-1 and SM3 algorithms. This indicates that while existing cryptographic libraries are effective at computing hash values for long messages, they are less suited for scenarios that require the computation of hash values for a large number of short messages. In such cases, a heterogeneous computing system is advantageous due to its greater computing resources. However, the performance improvement for MD5 is less significant, likely due to its simpler design and implementation compared to the other algorithms. Additionally, the execution time of the SHA-1 algorithm on Server A is 2.3 times faster than on Server B, highlighting our compiler’s ability to effectively utilize SHA-NI instructions and GPU resources to achieve optimal performance.

\subsection{Comparison with OpenCL}
\label{subsec:comp-opencl}
\begin{figure}[h]
\centering
\includegraphics[width=0.46\textwidth]{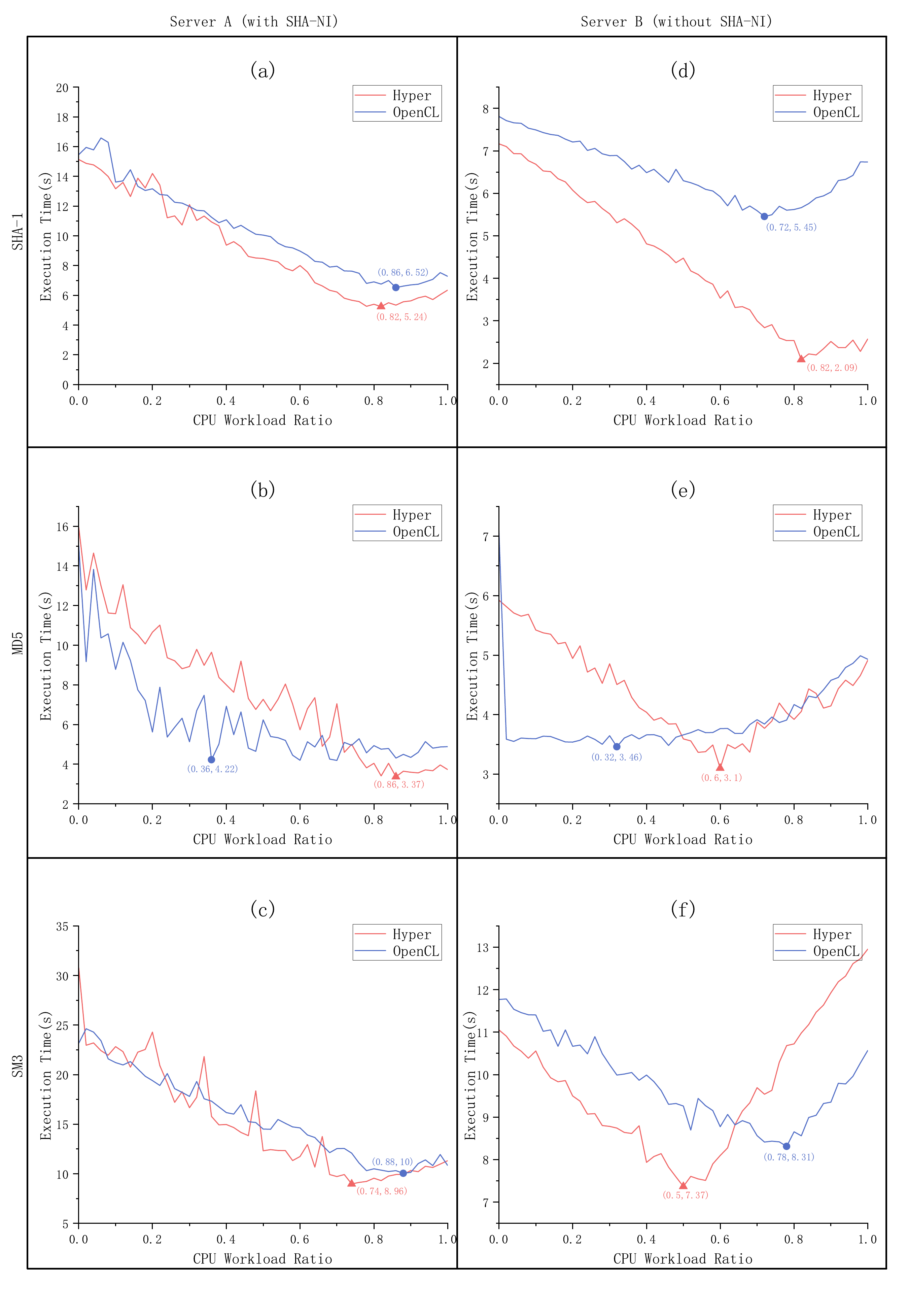}
\caption{Comparison results with OpenCL across different CPU and GPU workload ratios on different servers.}
\label{fig:compWithOCL}
\end{figure}

Figure~\ref{fig:compWithOCL} presents the comparison results between our \textit{hyper} dialect and OpenCL over a series of workload ratios within the same interval of 0.02. The results indicate that our approach consistently achieves better optimal performance than OpenCL under the same workload ratio search policy, with an average performance improvement of 1.93x for SHA-1, 1.18x for MD5, and 1.12x for SM3. Our method generally outperforms OpenCL in certain cases (Figure~\ref{fig:compWithOCL}(a), (c), and (d)), whereas in other instances (Figure~\ref{fig:compWithOCL}(b), (e), and (f)), OpenCL performs better under some workload ratios. These latter phenomena can be attributed to variations in hardware performance, leading to significantly different behaviors across systems despite using the same OpenCL kernel. However, our method still achieves superior performance in these cases, which can be attributed to reduced scheduling overhead. By implementing task splitting and parallel execution at the intermediate representation level rather than at the source language level, our approach provides finer granularity for applying specific optimizations. Additionally, although the performance gains may not appear substantial, our method avoids the complexity of OpenCL APIs by wrapping all parallel computing details within our prototype compiler, offering a more user-friendly solution.

\subsection{Device Scheduling Analysis}
\label{subsec:res-analysis}
In this subsection, we conduct additional experiments to delve deeper into the internal mechanisms underlying the optimal performance of our \textit{hyper} dialect, accompanied by a reasoned analysis.

In the scenario of the aforementioned experiment, two heterogeneous devices were utilized: an x86 CPU and an NVIDIA GPU. It is reasonable to assume that the execution time for both the CPU and GPU is linearly correlated with the increasing amount of data, given that this parallel computation task is free from data races, as each input message and each output hash value are stored in separate locations within an array. Consequently, under these conditions, we can formulate a straightforward hypothesis model:

\begin{align}
T_{cpu} =& \frac{N_{data} \cdot P_{cpu} \cdot x}{N_{core}} \label{eq:tcpu} \\
T_{gpu} =& \left(\frac{P_{gpu}}{N_{thread}} + \cdot (T_{alloc} + T_{memcpy})\right) \cdot N_{data} \cdot (1 - x) \nonumber \\
&+ O_{gpu} \label{eq:tgpu} \\
T_{opt} =& \max{(T_{cpu}, T_{gpu})} \label{eq:topt}
\end{align}

In Equation~\ref{eq:tcpu}, $T_{\text{cpu}}$ denotes the theoretical execution time on the CPU, given the total number of input data  $N_{\text{data}}$, the computational capability of a single CPU core  $P_{\text{cpu}}$ (representing the time required to compute the hash of a single message), the workload ratio allocated to the CPU $x$, and the number of available CPU cores  $N_{\text{core}}$. For the GPU, the theoretical execution time is detailed in Equation~\ref{eq:tgpu} and includes the memory allocation and deallocation time $T_{\text{alloc}}$, the data transfer time between the CPU and GPU $T_{\text{memcpy}}$, the time for computation per thread $\frac{P_{\text{gpu}}}{N_{\text{thread}}}$ , and a fixed overhead associated with invoking CUDA calls $O_{\text{gpu}}$. Thus, the optimal execution time $T_{\text{opt}}$ is determined as the maximum of $T_{\text{cpu}}$  and $T_{\text{gpu}}$, leading to a piecewise curve due to the linear characteristics of both $T_{\text{cpu}}$ and $T_{\text{gpu}}$, as described in Equation~\ref{eq:topt}.

\begin{figure}[h]
\centering
\includegraphics[width=0.46\textwidth]{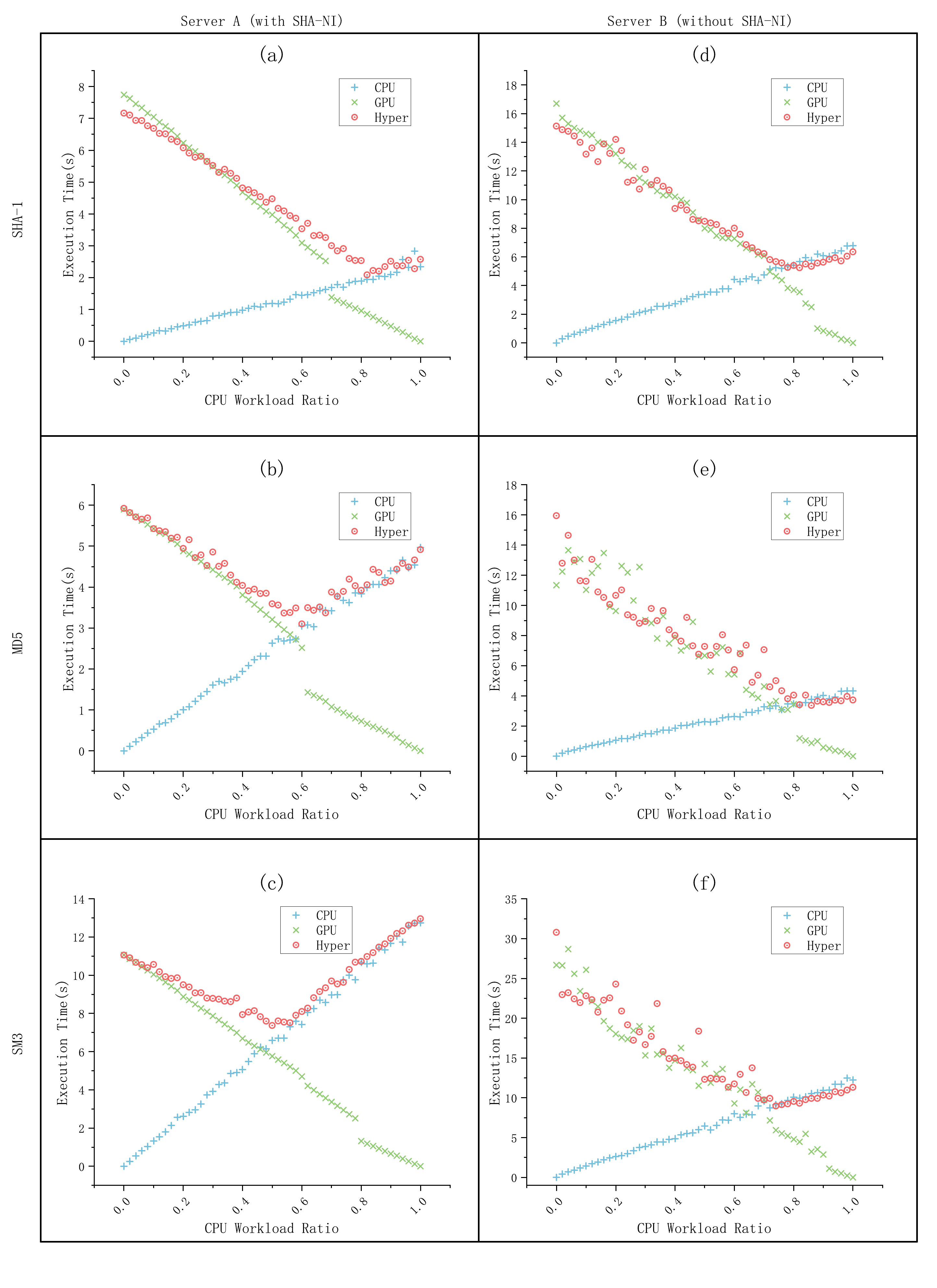}
\caption{CPU-only and GPU-only execution times across various input sizes, contrasted with hybrid execution times across different CPU and GPU workload ratios.}
\label{fig:dev-sche}
\end{figure}

To validate our hypothesis, we measured the actual execution times of CPU-only and GPU-only scenarios across varying numbers of messages, each with a fixed length of 9 bytes, on selected servers. The results are presented in Figure~\ref{fig:dev-sche}, which were generated by systematically varying workload ratios, as described in Sections~\ref{subsec:comp-openssl} and \ref{subsec:comp-opencl}, for Server A (which supports SHA-NI) and Server B (which lacks SHA-NI support). In Figure~\ref{fig:dev-sche}, the red “o” markers indicate the actual execution times of our approach for an input size of $9 \times 10^9$ bytes, allocated with varying workload ratios across different devices. Additionally, the blue “+” markers represent the CPU-only execution time with a fixed workload ratio of 0.0 for the GPU, while the green “x” markers indicate the GPU-only execution time with a fixed workload ratio of 0.0 for the CPU. The linear trends shown by these points partially support the linear models proposed for CPU and GPU execution times in Equations~\ref{eq:tcpu} and \ref{eq:tgpu}. Notably, the data points on either side of the inflection point of our measured execution times align well with segments of the curves for CPU-only and GPU-only scenarios, thereby substantiating our hypothesis as outlined in Equation~\ref{eq:topt}.

Additionally, Figure~\ref{fig:dev-sche} provides further insights into the performance dynamics. Firstly, the performance of SHA-1 in CPU-only mode on the server with SHA-NI support (Server A) is superior to that on the server without SHA-NI support (Server B), as indicated by the shallower slope of the blue points on Server A compared to Server B (see Figure~\ref{fig:dev-sche}(a) and Figure~\ref{fig:dev-sche}(b)). This demonstrates that optimal performance is achieved with a higher CPU workload ratio on Server A, highlighting how the optimal workload distribution varies across heterogeneous systems with different device capabilities. Secondly, each cluster of green points displays a “slope jump”, signifying a performance decline in GPU-only mode. This degradation is likely due to the implementation of a batching policy when memory requirements exceed the GPU’s capacity, leading to suboptimal GPU resource utilization. Improved performance could be realized if GPU computational power were fully utilized, even when data sizes exceed GPU memory limits, potentially resulting in a downward shift of the intersection point between the green and blue curves. Lastly, there is a noticeable gap between the theoretical optimal performance (the intersection of the green and blue curves) and the observed optimal performance (the inflection point of the red points). This discrepancy is likely due to inherent I/O and synchronization overheads, which could be further reduced with additional optimization.

It is important to note that the analysis presented here is limited to simple heterogeneous systems consisting of a single x86 CPU and a single NVIDIA GPU. Consequently, our model may not be directly applicable to more complex configurations involving multiple or varied devices. Nonetheless, this analysis highlights the significant potential of our \textit{hyper} dialect for effectively scheduling computing tasks across different devices. Future research will focus on developing more comprehensive scheduling models and expanding support to accommodate a wider range of hardware, which are essential components of our ongoing and future work.

\section{Conclusion \& Future Work}
\label{sec:conclusion}
In conclusion, this paper presents a novel approach to heterogeneous computing through the introduction of the \textit{hyper} dialect in the MLIR framework. By abstracting both data management and parallel computation across diverse hardware architectures, the \textit{hyper} dialect fills a critical gap in upstream MLIR and enables more efficient utilization of heterogeneous computing resources. The development of the \textit{HETOCompiler}, a cryptography-focused compiler prototype, demonstrates the practical applications of the \textit{hyper} dialect, providing significant performance improvements in typical cryptographic hash functions compared with OpenCL, highlighting the potential of the \textit{hyper} dialect to enhance computational efficiency in diverse computing environments. In the future, we plan to extend our work's compatibility to include more platforms and focus on ML-based workload scheduling algorithms for more complex parallel computing scenarios.

\bibliographystyle{IEEEtran}
\bibliography{mybib}

\end{document}